# Black Boxes, White Noise: Similarity Detection for Neural Functions

Farima Farmahinifarahani[a] and Cristina V. Lopes[a]

a    University of California, Irvine, USA

**Abstract**    Similarity, or clone, detection has important applications in copyright violation, software theft, code search, and the detection of malicious components. There is now a good number of open source and proprietary clone detectors for programs written in traditional programming languages. However, the increasing adoption of deep learning models in software poses a challenge to these tools: these models implement functions that are inscrutable black boxes. As more software includes these DNN functions, new techniques are needed in order to assess the similarity between deep learning components of software.

    Previous work has unveiled techniques for comparing the representations learned at various layers of deep neural network models by feeding canonical inputs to the models. Our goal is to be able to compare DNN functions when canonical inputs are not available – because they may not be in many application scenarios. The challenge, then, is to generate appropriate inputs and to identify a metric that, for those inputs, is capable of representing the degree of functional similarity between two comparable DNN functions.

    Our approach uses random input with values between −1 and 1, in a shape that is compatible with what the DNN models expect. We then compare the outputs by performing correlation analysis.

    Our study shows how it is possible to perform similarity analysis even in the absence of meaningful canonical inputs. The response to random inputs of two comparable DNN functions exposes those functions' similarity, or lack thereof. Of all the metrics tried, we find that Spearman's rank correlation coefficient is the most powerful and versatile, although in special cases other methods and metrics are more expressive.

    We present a systematic empirical study comparing the effectiveness of several similarity metrics using a dataset of 56,355 classifiers collected from GitHub. This is accompanied by a sensitivity analysis that reveals how certain models' training related properties affect the effectiveness of the similarity metrics.

    To the best of our knowledge, this is the first work that shows how similarity of DNN functions can be detected by using random inputs. Our study of correlation metrics, and the identification of Spearman correlation coefficient as the most powerful among them for this purpose, establishes a complete and practical method for DNN clone detection that can be used in the design of new tools. It may also serve as inspiration for other program analysis tasks whose approaches break in the presence of DNN components.

## The Art, Science, and Engineering of Programming

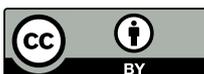

**Black Boxes, White Noise**

## 1 Introduction

An increasing number of software systems, both open source and proprietary, include "AI components" that consist of deep neural network (DNN) models implementing specific input/output functions [4, 7, 18, 30, 38, 40, 41]. These DNN functions are inscrutable black boxes, due to them being matrices of numbers: their structure does not disclose any insights on their input/output behavior (see Figure 1). Even by having access to models' files and being able to inspect their architecture and learned weights, it is not possible to gain insights on models' functions by inspecting this information. This brings new challenges to the production and maintenance of software systems [1], resulting in the need to extend conventional software engineering practices and tools with novel techniques that can be used for neural network components.

Finding similarities in code, known as code clone detection, has been a focus of software engineering in the past few decades [9, 31, 32]. Traditional code clone detection techniques and tools, however, are not prepared for DNN functions. The function of a DNN model comes largely from its training data [1]; given different data, similar setup scripts and network architectures may end up producing completely different models [35], and, hence, completely different functions. Therefore, functional comparison of DNN models cannot rely on syntactic or structural properties of the models themselves. Instead, it must compare the functions' outputs on canonical inputs, akin to testing. Given a canonical set of test inputs, when the outputs of two functions are sufficiently similar, then they are similar.

This intuitive statement about DNN function similarity, however, needs to be operationalized in two fronts: (1) what inputs should be used to assess similarity, and (2) how can we quantify "sufficiently similar"? With respect to (1), test sets are the obvious candidates, but they may not always be available. In fact, in some cases we may not even know what the models are supposed to do, much less what data was used to train and test them. With respect to (2), methods and metrics that have been used for comparing traditional code, such as functional equivalence [15], fall short of capturing similarity of DNNs. Recent work in this area focuses on statistical methods and metrics to compare the representations learned by DNNs at various layers, using meaningful inputs. One promising family of methods is the Canonical Correlation Analysis (CCA) [13], which has been recently applied to neural networks [23, 28]. Another promising, but more constrained, metric is the Spearman rank correlation, which has been applied in a small study using meaningful canonical inputs [34].

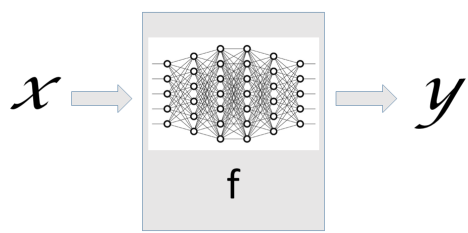

■ **Figure 1** DNN function.

Finally, for classifiers in particular, we can use a simple overlap metric that quantifies how many times two different models agree on the classification of the same inputs.

Our paper presents a study of methods and metrics for DNN similarity detection in the absence of meaningful canonical inputs, using a dataset of 56,355 classifiers collected from GitHub. Our findings indicate that random inputs are perfectly suitable for similarity detection. The three metrics we studied produce very good results, all of them above 76 % accuracy in the detection of similar/dissimilar functions. Spearman rank correlation stands out as the most accurate within the scope of our study, with 94 % accuracy. CCA is the most general metric, even more than Spearman correlation, as it can be used to compare any two functions with any number of outputs; but within our study, CCA has the highest number of erroneous similarity classifications. The simple overlap metric has the highest uncertainty, but it is able to detect similarities that the other two metrics failed to detect.

The key contributions of this work are as follows:

- We formulate the problem of finding functional clones of DNN models, and operationalize DNN function similarity detection. To the best of our knowledge, this is the first large study of this problem.
- We show that random inputs can be used in lieu of canonical inputs for the purposes of finding similar classifiers. This finding enables and simplifies the implementation of model clone detectors, as model-specific test data becomes unnecessary.
- We unveil the strengths and weaknesses of three similarity metrics, being that Spearman rank correlation stands out as the most accurate for classifiers.
- We curate a dataset of more than 56,000 compiled Keras [6] classifiers from GitHub, clustered based on their input and output shapes. This dataset can be used not just for reproducing our work but also for further analysis of DNN models.

The rest of this paper is structured as follows. We discuss the details of the problem of DNN functions similarity detection in Section 2, followed by a detailed description of the design of this study in Section 3. The experiments are detailed in Section 4, and a sensitivity analysis is presented in Section 5, followed by related work in Section 6, threats to validity in Section 7, and conclusions and future work in Section 8.

## 2 Problem Definition

Here, we first explain the need to investigate processes and metrics for finding similar DNN functions in the absence of access to their training/test data. We then define the concept of function similarity and how it can be quantified.

### 2.1 Motivation

**Model Theft.** As reported in the literature [2, 39], software theft is a pervasive problem that costs billions of dollars each year. One common scenario of software theft is when employees leave companies and take copies of the proprietary software code or data, a scenario that is on the rise at an alarming rate [14]. With recent advances in AI and machine learning, and the replacement of traditional software components with DNN models, software is not just computer instructions anymore. Reports indicate that model theft is starting to occur, too, both by directly copying the model file that may be slightly modified later [21] and by duplication of the model's functionality using model extraction attacks [16, 29, 36]. In any of these cases, theft investigation requires the examination and comparison of DNN functions without their training or test data – the training data is not packaged for deployment, and may not be available for analysis for a variety of reasons including but not limited to privacy or security concerns. In fact, when presented with two large systems for comparison, or a comparison of one system against a large base of software artifacts, an investigator of software theft may often be a third-party who does not even know what kind of functions the included models implement. Investigation of functions similarity is particularly useful in the detection of model theft since when a model is stolen, it is its function that is of interest, and therefore, the ultimate function of the model is expected to remain similar.

**Malware in App Stores.** Another application of DNN functions clone detection is in app stores that scan the apps for malicious components. The number of mobile apps that use DNN models to support their main features is on the rise: a study in 2019 [44] showed that 81 % of the apps that included DNN models, used these models to support their core features. Traditional program analysis will soon be insufficient to detect malicious behavior, and DNN functional comparison will need to be performed without needing to have access to the training and test data (the apps are not submitted with their data). Using the methods and metrics investigated in our study, when apps include DNN functions, those functions can be automatically compared against a large base of known functions previously identified in malicious apps.

**Model Search.** Another use-case of DNN functional comparison is in *model search*. An example of this is in searching for a DNN function that is similar to a model at hand, but has a better accuracy or generalization ability. DNN functional comparison can be used to compare the model at hand's functionality with the models available in a repository of trained models, and producing similarity scores for each model pair. This scenario can, for example, happen when searching for a pre-trained model and customizing it to one's specific needs by performing further training over it (similar to transfer learning). The benefits of further training of an already trained model (instead of starting from scratch) are easing the development of DNN functions for AI-enabled software, reducing the time and computational costs of model development and training, increasing the performance and capabilities of the final model, and overcoming insufficient training data.

### 2.2 Functional Similarity

Functional similarity measures the extent to which the functions are similar. In here, we should clarify our use of the words *function* and *model*. A *function* is a mathematical object described by its input/output behavior, independently of how it is implemented; a *model* is a particular implementation of a function, one that uses neural networks; this is when we refer to functions as *DNN functions*. Unlike traditional code, comparing the training and testing scripts of DNN models is useless for the purposes of assessing their functions' similarity since the functionality of models is mostly based on the data that is used for training them; it is common to have identical training scripts that produce entirely different input/output behaviors. Therefore, detecting functional similarity of DNN models requires actual input/output analysis.

Before defining functional similarity, we define functional *equivalence*. In traditional code, functional equivalence has previously been defined as the equivalence of output given the same input accounting for permutations of inputs and outputs [15]. We generalize that definition for DNNs. Let $f_1$ and $f_2$ be two functions; let $I$ be a set of inputs for $f_1$. We say that $f_2$ is functionally *equivalent* to $f_1$ if $f_2(\phi(i)) = f_1(i), \forall i \in I$, where $\phi(i)$ is a geometric transformation [3] of the input $i$, which could be the identity map, but also scaling, reflection, projection, permutation and other such transformations. The geometric transformation of the inputs is a generalization of the permutations of the input used in [15]. The intuition for both is that similarities can be searched for, and found, even when the inputs are not exactly the same, but are related via these transformations. For example, a digit recognition function that takes 28×28 inputs in the range of -1 and 1 can be compared to another function that takes 28×28 inputs in the range of 0 and 255 (this is an example of scaling).

Functional equivalence, however, is of limited interest. Functions can be very similar without producing the exact same outputs for the same (or transformed) inputs. Consider, for example, two functions, each with one single numerical output, which, for the same inputs produce outputs such that $o_2 = 2o_1, \forall o \in O$ – i.e. the second function's output values are always double the first function's output values; clearly, these functions are doing something very similar even if the values are scaled by a factor of 2. As a second example, consider two functions which, for the same 10,000 inputs, produce the exact same outputs in 9,000 cases, and different outputs in 1,000 cases; again, these two functions are similar, even if their outputs differ 10% of the time.

Given this, similarity of functions is best captured by measuring the *correlation* between the outputs [23, 28, 34]. If for the same (or transformed) inputs, the outputs of two functions are correlated, then they are similar. Formally, $f_1$ and $f_2$ are similar if

$$Corr(f_1(i), f_2(\phi(i))) > \theta, \forall i \in I \qquad (1)$$

where *Corr* is a correlation metric, $\phi$ is a geometric transformation, and $\theta$ is a similarity threshold.

In the case of functions with more than one output, and even with different number or ordering of outputs, the correlation metric must reflect an aggregation of the several outputs using some aggregation method. In this general definition, *Corr* stands for the aggregated correlation of the outputs.

## 3  Study Design

In this section, we elaborate on the goals and scope of our work, followed by a description for the datasets used in our experiments.

### 3.1  Goals and Scope of the Study

The definition of functional similarity given in Equation 1 requires the existence of inputs for the purpose of input/output analysis. In general, meaningful canonical inputs, such as training or test data, may not be available. As such, the goal of this study is twofold: (1) we want to investigate whether random inputs can be used, instead of canonical ones, to detect DNN functions similarity; and (2) we want to understand the trade-offs of different similarity metrics. Our findings shed light on the process of DNN functional similarity detection and on the applicability of individual metrics in different scenarios.

Equation 1 is generic, in the sense that it can be used to compare any two DNN functions independent of their nature and input/output shapes and dimensions. For the purposes of this study, we limit our attention only to classifiers; moreover, we narrow the scope of this study to single-label multi-class DNN classifiers with compatible input shapes and the same number, and ordering, of output classes. Compatible input shapes here means that the input shapes of $f_1$ and $f_2$ can be derived from each other by simple reshape transformations. Examples include transforming $28x28$ 2-dimensional inputs into 784 1-dimensional inputs, and vice-versa. Excluded from our study are classifiers whose domains have more general geometric transformations such as image resizing (e.g. $256x256$ vs. $128x128$). Also, out of scope are classifiers with different number of output classes, such as one with 10 classes and another with 11. The input shape and output shape compatibility here is checked by inspecting models' files; therefore, we assume having access to models' files.

These constraints were followed simply to tame the complexity of the empirical analysis, as trying to compare any two models would not only increase the dataset considerably but would also expand the results and analysis beyond the page limit.

Our study is in two parts: the first part (Section 4) has two goals: we study if random inputs are suitable for DNNs functional similarity detection (Goal 1), and what are the trade-offs in using different similarity metrics (Goal 2). Studying Goal 1 is important because DNN models' are trained on canonical inputs and are supposed to produce output on such inputs. When canonical inputs are substituted with random inputs, such inputs can be white noise and the results generated by DNNs might be random predictions that are not able to capture the correlations among similar models. As an example, the image in Figure 2 shows a $28x28$ picture generated using random inputs. If we feed this picture to a DNN function trained to distinguish hand written digits, the model's output could be a random value that may not be reliable for capturing the similarities and dissimilarities. Therefore, it is important to study whether random inputs can be useful for the purpose of similarity detection. We do this by comparing the models' accuracy (the fraction of times models' predicted labels match the ground truth labels) on 5 well-known classification datasets (which establishes the ground truth of functionality) against their similarity with reference models, that are trained on the target well-known datasets, calculated using random-inputs. Related to Goal 1, we also study if it is possible to use random-inputs for detecting similarity within models for which we have no information. We investigate this since normally, the person who is performing the comparisons does not have any knowledge about the two models being compared. We therefore, run experiments using models that are unknown to us to assess the feasibility of DNN similarity detection using random inputs in such scenarios. Here, a manual inspection of models' GitHub repositories helps us in reasoning about the predicted similarity values.

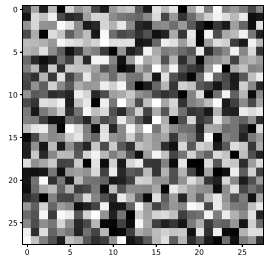

**Figure 2** Example of an image generated using random inputs.

To explore Goal 2, we compare three similarity metrics with respect to recall, precision, and accuracy in a set of experiments using well-known datasets.

Specifically, we investigate the following research questions:

- RQ1: Is it possible to detect similar and dissimilar DNN functions using random inputs?
- RQ2: How do various similarity metrics compare?
- RQ3: Is it possible to detect similarities within unknown models using random inputs?

Second (Section 5), we perform a sensitivity analysis where the goal is to understand how the randomized nature of inputs and a set of models' training related properties affect the similarity results. As such, we investigate the following additional questions:

- RQ4: What is the effect of input generation randomization on the functions' response to similarity detection?
- RQ5: What is the effect of the models' training related characteristics (such as training randomization, accuracy, architectures, training datasets) on their response to similarity detection?

## 3.2 Evaluation Datasets

The datasets used in our experiments are as follows.

### 3.2.1 Main Experiments Dataset (D56K)

This dataset consists of **56,355** classifiers collected in the following manner. We used the GitHub code search API[1] to look for files with *.h5* extension, a widely used extension for saved Keras [6] models.[2] In total, we obtained 340,933 h5 files, of which we were able to download 335,789. Next, we attempted to load these models into the Keras environment to separate DNN models from other possible objects saved in h5 files, and we identified 102,602 DNN models. After loading the models, we clustered them based on their input and output shapes, which was obtained by inspecting models' files. Finally, since the focus of our work is on classifiers, we only retained models whose output layer has either a *sigmoid* (to fetch binary classification functions) or a *softmax* (to fetch multi-class classification funcions) activation function. This resulted in a total of 56,355 models clustered into 6,431 groups.

### 3.2.2 Sensitivity Analysis Dataset (D14)

For the experiments related to RQ4 and RQ5, we trained 14 models using MNIST (10 class hand-written black digit on white background classification) and FMNIST (10 class pieces of clothing classification), two popular and publicly available datasets that have extensively been used in previous research [25, 27, 43]. These models are presented in Table 1, and are grouped in clusters such that models in each cluster have one varying property with respect to the first model, MN Ver1, which we consider to be the reference model. The group "Instances" consists of two identical models in all aspects, except for being trained separately, therefore being subjected to the randomness of the training process. The next group, "Train Params", consists of models with different training parameters as to the reference model, the "Architectures" group includes models that have architectural differences with respect to the reference model, and the "Datasets" group consists of models that have been trained on different datasets, with varying levels of similarity compared to the training data of the reference model. The column "Acc" shows the accuracy of models on their own test data, and the column "X-Acc" is the accuracy of the models on the reference model's test data. X-Acc serves as the ground truth for similarity of the models with respect to the reference model. Therefore, MN Rev Color (which has a reverse coloring of background and foreground as to the reference model) and FMN are considered to be dissimilar to the reference model as they both have a very low X-acc.

### 3.3 Functional Similarity Metrics

To assess the functional similarity between two classifiers, we analyze their predictions on the same random inputs. In order to quantify similarity, we consider three metrics: Canonical Correlation Analysis (CCA), Spearman rank correlation ($\rho$), and a simple overlap metric. The choice of these metrics is based on previous similarity

---

[1] https://docs.github.com/en/rest/reference/search#search-code. Accessed: 2023-02-03
[2] There are other external storage formats as well, but we limited our search to h5, because of its simplicity.

■ **Table 1** Sensitivity analysis: trained models (D14 dataset)

| Group | Name | Description | Acc | X-Acc |
|---|---|---|---|---|
| Instances | MN Ver1 (Ref) | 2 layers fully connected with ReLU in 1st layer, #epochs=10, learning rate =0.001, MNIST training data scaled to (0,1) | 98.16% | 98.16% |
|  | MN Ver2 | Same architecture, training params, and training data as Ref model, trained separately | 98.15% | 98.15% |
| Train Params | MN Long Tr | Same architecture, training data, and learning rate as Ref model, trained for 50 epochs | 98.35% | 98.35% |
|  | MN LR01 | Same architecture, training data, and epochs as Ref model, learning rate=0.01 | 97.80% | 97.80% |
|  | MN LR18 | Same architecture, training data, and epochs as Ref model, learning rate=0.18 | 75.92% | 75.92% |
| Architectures | MN Sig | 2 layers fully connected with Sigmoid in 1st layer, training parameters and training data same as Ref model | 97.68% | 97.68% |
|  | MN Deep NN | 4 layer fully connected with ReLU in the 1st 3 layers, training parameters and training data same as Ref model | 98.21% | 98.21% |
|  | MN CNN | CNN with 2 convolution, 2 pooling, and 2 fully-connected layers, training params and training data similar to Ref model | 99.18% | 99.18% |
|  | MN CNN2 | Same architecture, training params, training data as MN CNN, trained separately | 99.06% | 99.06% |
| Datasets | MN 1st Batch | Same architecture and training params as Ref model, trained on MNIST training data's 1st 30K rows in the scale of (0,1) | 97.55% | 97.55% |
|  | MN 2nd Batch | Same architecture and training params as Ref model, trained on MNIST training data's 2nd 30K rows in the scale of (0,1) | 98.39% | 98.39% |
|  | MN No-scale | Same architecture and training params as Ref model, trained on MNIST training data in range of (0,255) | 97.21% | 90.74% |
|  | MN Rev Color | Same Architecture and training params as Ref model, trained on MNIST black background & white digit scaled to (0,1) | 95.27% | 1.67% |
|  | FMN Model | Same architecture and training params as Ref model, trained on Fashion MNIST training data scaled to (0,1) | 88.88% | 8.35% |

studies: CCA has been used for measuring the similarity between neural network representations [23, 28], and Spearman correlation is a suitable similarity metric [5] that has been used in assessing models' functional similarity when canonical inputs are available [34]. The overlap metric is inspired by the study of code functional equivalence by Jiang et al. [15], adapted to measure similarity; here we assume that the order of the outputs is the same, and simply measure the degree of overlap on classifiers' outputs given the same inputs. Details of each of these metrics follow below.

CCA is a statistical method for inferring the relationship between two set of variables, by finding linear relationships between them such that the correlation between the linear relationships is maximized [23, 34]. CCA has previously been used in measuring the similarity between the learned representations of different layers of different neural networks by feeding canonical datasets to the networks and measuring CCA on the activations of these layers [23, 28]. CCA is very generic, and is capable of finding the relationship between two sets of variables with different cardinalities, making it a viable choice in assessing the functional similarity where the functions' output dimensions are different [34].[3] The correlation coefficient in CCA ranges from 0 to 1 [33]. We calculated the CCA correlation coefficients over the output prediction probability vectors of a pair of classifiers over the inputs fed to them, which produced $n$ correlation values where $n$ is the number of output labels. We then aggregated the results by calculating the *mean* value over the $n$ correlation values, similar to the method explained in [23].

Spearman correlation [34] is a non-parametric correlation metric [5] which has been used in previous studies of neural network similarity using canonical inputs [34]. Apart from this, what makes Spearman a suitable choice is that by doing rank correlation, this metric is not limited to measuring the linear relationship between the variables. Since various data scalings may happen on the training data of DNNs, this property makes this metric capable of capturing similarity even when non-linear data transformations happen on data (specifically, on training data's output values). Moreover, in a general case, if the monotonicity over the same inputs of functions is similar, then the functions are also similar. Spearman correlation ranges between $-1$ and $1$, where positive values denote a direct correlation, negative values denote an inverse correlation, and values close to zero (typically between $-0.1$ and $0.1$) denote the absence of correlation. To perform Spearman correlation on the outputs of classifiers, we considered the probability values reported for each output label as one variable and calculated the correlation values for this output across all $m$ samples. In other words, if function $f_1$ has n output classes $\{c_1^1, c_1^2, .., c_1^n\}$ and function $f_2$ also has n output classes $\{c_2^1, c_2^2, .., c_2^n\}$, we calculate n correlation values $\{\rho^1, \rho^2, .., \rho^n\}$ such that $\rho^i = Correlation(c_1^i, c_2^i)$. To aggregate the results, we take the *mean* of these $n$ correlation values and use that as the final similarity value. Using this metric, therefore, requires the two functions to have the same output shapes and dimensions, and to interpret them in the same order, which, as explained before, is a simplifying assumption of our study.

Finally, the last and the simplest similarity metric that we used is a simple *overlap*, i.e. the number of times the two functions agree on the classification of the random inputs. This metric simply measures the percentage of the times that the output value of two classifiers are the same. It is very similar to the accuracy metric: if the predictions made by the reference classifier are considered as the ground truth, then this metric calculates how accurate the candidate model is with respect to the reference model. Given its categorical nature, this metric is only applicable to classifiers.

In summary: CCA is the most generic metric, as it can be used to compare any two functions; our use of Spearman correlation assumes the same number and order of outputs; and the simple overlap metric assumes that functions perform classification.

### 3.4 Similarity Thresholds

The similarity metrics need to have a *threshold* at which we classify functions as similar or dissimilar. For CCA and Spearman, we can tune the threshold empirically using the general and accepted guidelines for correlation ranges: a value between 0 and 0.1 is generally considered no correlation, a value between 0.1 and 0.2 is considered weak positive correlation, between 0.2 and 0.5 is moderate and positive, and more than 0.5 is a high positive correlation. For the purpose of similarity, we consider moderate and above correlation values (values>= 0.2) as similar, weak correlation values (0.1 <values< 0.2) as uncertain, and values<= 0.1 as dissimilar. We performed various sensitivity analyses in our experiments, and observed that these thresholds work well in capturing similar functions while minimizing the number of false predictions. This means that for Spearman that reports inverse correlations with negative values, these values are considered as dissimilar as well, since an inverse relationship also shows differences between the predictions.

---

[3] In the case of our study, and as explained before, the output dimensions of comparable models are never different.

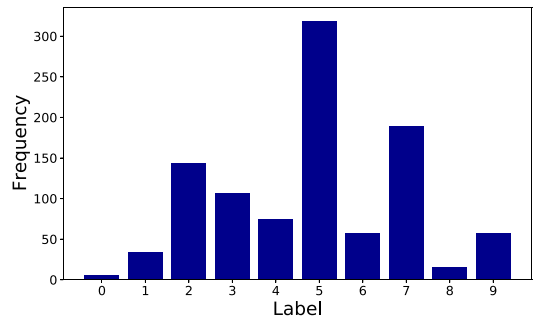

**Figure 3** Frequency of Labels with Unconstrained Random Inputs

For the overlap metric, the calculated similarity values show the fraction of times two classifiers predict the same label over the same input. This metric is similar to accuracy where the number of times the predicted labels match the ground truth labels is counted; in our case, the ground truth is the reference model's predictions and we count the number of times the candidate model's predicted labels match those of the reference model. One important note here is that accuracy has been shown to be negatively affected if the underlying input dataset has an unbalanced distribution of labels [22]. This problem is extended to the Overlap metric as well: when the reference model's predicted labels over the inputs have an unbalanced distribution, the calculated Overlap values become unreliable and it will be impossible to specify a similarity/dissimilarity threshold that works for all kinds of classifiers, with any number of labels. Based on our experiments, it is very likely that over the unconstrained random inputs, models predict some labels much more than the other ones (due to model biases, for example). As an example, Figure 3 shows the frequency of labels predicted by a 10 class digit recognition model over unconstrained random inputs. As the plot shows, label 5 has been predicted much more than the other labels while for example, labels 0 and 8 appear very few times. This issue is explained in more depth in the appendix (Section A).

To address the above issue with unconstrained random inputs, we generate and use balanced random inputs with the Overlap metric (the process for generating such inputs is explained in the appendix). In particular, if the input dataset $d$ has $m$ rows and is balanced with respect to the reference model $m_r$ which has $n_r$ output classes

**Black Boxes, White Noise**

(that is $m_r$ predicting each output class for $\frac{m}{n_r}$ times on $d$), then any arbitrary model (that also has $n_r$ output classes), just by pure chance, can agree with $m_r$ for $1/n_r$ times. Therefore, when the level of agreements between $m_r$ and a model on $d$ is at the level of chance ($\lesssim 1/n_r$), the two models cannot be classified as similar. If the agreements level is much higher than $1/n_r$, for example twice the level of chance ($2/n_r$), then something other than chance is at play and is a strong indicator that the models are doing something similar. The upper threshold of $2/n_r$, however, can be too strict given model's imperfect accuracies. In the case of binary classifiers, for example, this threshold translates to 100 % similarity which does not happen in practice. Even the most similar models do not agree with each other 100 % of times, and their accuracies affect their agreements level. In general, we do not know the models' accuracies, but we know that a 100 % accuracy is rare. Therefore, the similarity threshold $2/n_r$ should be multiplied by an empirical factor $\alpha$, and become $2\alpha/n_r$. For most good models, an accuracy above 90 % is expected and therefore, depending on whether both models' accuracies are taken into account or only one of them, $\alpha$ can have a value between 0.81 and 0.9. In our experiments, we observed that $\alpha = 0.9$ works well while still eliminating the false positives.

Based on this, when using the Overlap metric, we generate *balanced random inputs* with respect to the reference model, where with $n_r$ number of output classes for both models, a similarity $<= 1/n_r$ is considered dissimilar, a similarity $>= 2\alpha/n_r$ is considered similar, and between these two is uncertain where we cannot comment on models' similarity with confidence.

### 3.5 Similarity Inspection Pipeline

The pipeline that we designed to carry out this study, consists of two high-level modules: (i) a compatibility verification module and, (ii) a functional similarity module. We designed this pipeline in a way that it can be used for comparing any two classifiers, and not just the models inside one cluster in our datasets. In compatibility verification module, the models' output compatibility is verified by checking that the two models have the same output shapes. This includes verifying that the two models predict the same number of classes and that they have the same activation function in their output layers. This is done by inspecting the architecture of the model which is available through its file. Input compatibility can go beyond exact input shapes: models can be input compatible if their input shapes can be derived from one another. This is to accommodate various dataset reshapings that engineers may do when training the models. Therefore, for the input shape compatibility verification, given $i_r$ and $i_c$ as the input shapes for the reference and the candidate models respectively, we flatten the input shapes and compare them: two models have compatible input iff $length(flat(i_r)) == length(flat(i_c))$.

Finally, functional similarity is investigated by generating unconstrained random inputs (float values between -1 and 1) for CCA and Spearman, and balanced random inputs (with respect to reference model's predictions) for the Overlap metric, and then measuring the similarity on the generated inputs using each metric. The implementation of this pipeline is available at *this link*.

■ **Table 2** Similarity metric' successes and errors

| Query Model | Similarity Range | | | | | | Dissimilarity Range | | | | | | Uncertain. Range | | |
|---|---|---|---|---|---|---|---|---|---|---|---|---|---|---|---|
| | #TP | | | #FP | | | #TN | | | #FN | | | #Total | | |
| | CCA | Spr | Ovl | CCA | Spr | Ovl | CCA | Spr | Ovl | CCA | Spr | Ovl | CCA | Spr | Ovl |
| MN same shp | 60 | 72 | 68 | 0 | 0 | 0 | 23 | 23 | 21 | 4 | 2 | 2 | 12 | 2 | 8 |
| MN RevClr same shp | 1 | 1 | 1 | 6 | 0 | 0 | 40 | 99 | 97 | 0 | 0 | 0 | 53 | 0 | 2 |
| FMN same shp | 11 | 15 | 15 | 0 | 0 | 0 | 82 | 82 | 41 | 0 | 0 | 0 | 4 | 0 | 41 |
| Iris | 339 | 336 | 272 | 13 | 2 | 0 | 12 | 27 | 23 | 0 | 1 | 1 | 7 | 5 | 75 |
| Sonar | 60 | 60 | 25 | 0 | 0 | 0 | 1 | 1 | 1 | 0 | 0 | 0 | 1 | 1 | 36 |
| MN compat shp | 34 | 39 | 37 | 0 | 0 | 1 | 59 | 58 | 49 | 3 | 1 | 2 | 4 | 2 | 11 |
| FMN compat shp | 20 | 39 | 38 | 0 | 1 | 2 | 52 | 51 | 29 | 2 | 1 | 1 | 20 | 2 | 24 |
| CNN compat shp | 0 | 0 | 12 | 0 | 0 | 1 | 59 | 59 | 48 | 41 | 38 | 5 | 0 | 3 | 34 |
| **Total** | **525** | **562** | **468** | **19** | **3** | **4** | **328** | **400** | **309** | **50** | **43** | **11** | **101** | **15** | **231** |

# 4 Experiments

The details of the experiments conducted to answer our research questions are detailed in the rest of this section.

## 4.1 Accuracy vs. Functional Similarity

Here we answer RQ1 and RQ2. We selected 5 known datasets – *MNIST, Fashion MNIST, MNIST Reverse Color, Iris [10], and Sonar [8, 11]* – and trained classifiers on them to serve as our reference models (from these datasets, the first three ones classify images, as described in detail in Section 3.2.2, and the latter two contain numerical features describing types of Iris plants in case of the Iris dataset and numberical features describing sonar signals in case of the Sonar dataset). We then identified the clusters within our D56K dataset whose input and output shapes match, or are input compatible with, those of these models, and measured the accuracy of the models fetched from the clusters on the corresponding canonical test datasets; i.e., the fraction of times models' outputs agree with the labels in the canonical test datasets. By using the canonical test sets, we are able to have the ground truth of the functionality of the models through their accuracy values: empirically, we consider that when models have accuracy above 65 % on a canonical test dataset, they implement a good-enough classification of that dataset's inputs. When the accuracy falls below 50 %, we consider the models to be classifying something different, and in between 50 % and 65 % accuracy, we consider the models to be undecided, with just a vague functional similarity. To make sure we calculate the correct accuracy for the models, where applicable, we applied various common feature scaling techniques to the inputs, measured the accuracy of each model on all the scaled and non-scaled datasets, and considered the maximum accuracy value among these as the true accuracy of the model. In the end, 16 cases fell in the undecided region; we excluded those from the calculations of successes and errors (Table 2, presented later) and recall, precision, and accuracy computations (Table 3, also presented later).

Having established the ground truth of whether the models perform the chosen test sets' classifications or not, we then fed random inputs to both the reference models and these models, and measured the similarities using the metrics described in Section 3.3.

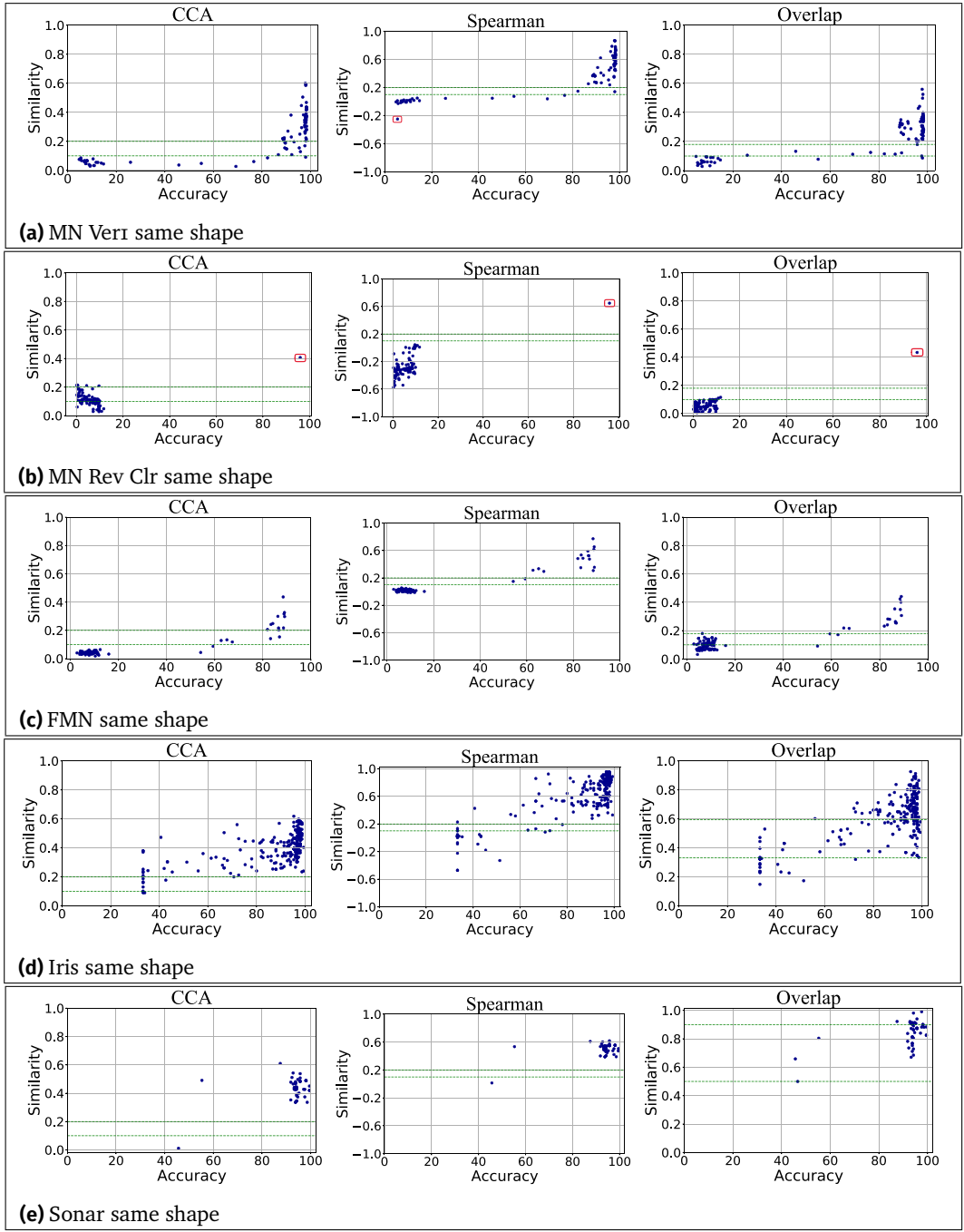

**Figure 4** Accuracy vs. similarity for Same Shape GitHub models

**Table 3** Precision, recall, accuracy per similarity metric

| Precision | | | Recall | | | Accuracy | | |
| --- | --- | --- | --- | --- | --- | --- | --- | --- |
| CCA | Spr | Ovl | CCA | Spr | Ovl | CCA | Spr | Ovl |
| 96 % | **99 %** | **99 %** | 85 % | **91 %** | 76 % | 83 % | **94 %** | 76 % |

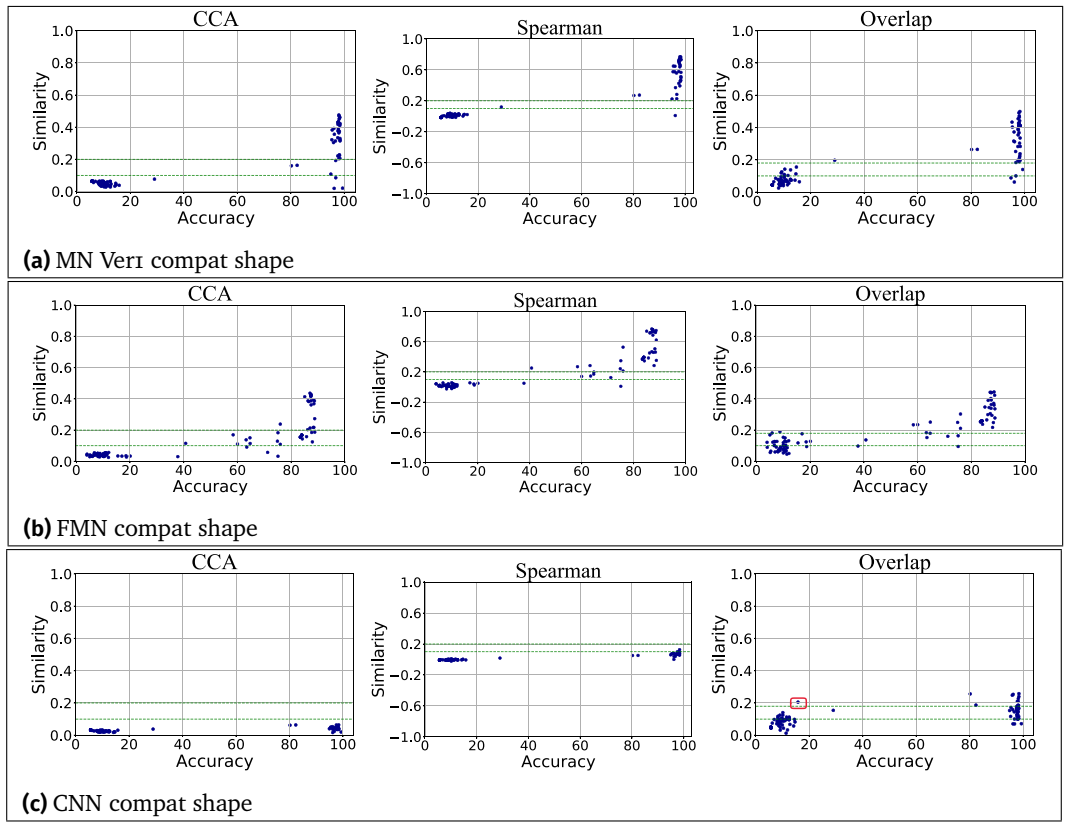

**Figure 5** Accuracy vs. similarity for Compatible Shape GitHub models

For each reference model and each metric, we show scatter plots that depict the relationship between the candidates models' (models being compared with the corresponding reference model) accuracy (x-axis), taken as the ground truth, and their similarity score with the reference model using random inputs (y-axis). These plots are shown in Figure 4 (same shape models) and Figure 5 (compatible shape models). The dots in these plots denote the *similarity* vs *accuracy* for each candidate model when being compared with the reference model addressed in the plot's caption. For example, in Figure 4(a), each dot shows this relationship between each candidate model being compared to the model named *MN Ver1*. In these plots, "same shape" means that the reference and its candidates have the exact same input shapes, and "compat shape" means that input shapes are not exactly the same, but compatible. The red boxes around each dot demonstrate interesting cases which are discussed later at the end of this section. The green dashed horizontal lines on each plot show the dissimilarity and similarity thresholds for the corresponding similarity metric (in case of overlap metric, the thresholds vary based on the number of output classes, $n$). The region below the bottom green dashed line shows the dissimilarity region, the region between the two lines shows the uncertainty region (where based on the thresholds, the metric cannot comment about models' similarity with high certainty), and the region above the top green dashed line shows the similarity region. An ideal metric would have the least number of points in the uncertainty region.

As a more detailed example, Figure 4(d) shows the accuracy vs. similarity between our trained Iris reference model and 376 models we found with the exact same input shape (a vector of size 4) and output shape (a vector of size 3); the accuracy values come from testing those real-world models with the canonical Iris dataset, and the similarity values for each plot come from comparing these models using the corresponding metric and random inputs. The green threshold lines are at 0.1 and 0.2 for CCA and Spearman, and at 1/3 (chance) and $2 \times 0.9/3$ (not chance) for Overlap, given 3 output classes.

In total $1,037$ comparisons were done per metric, therefore, there are $3,111$ datapoints shown in all of the plots (we have 3 metrics, therefore, $1,037 * 3 = 3,111$).

Additionally, Table 2 quantifies each metric's successes and errors for all of these comparisons. The columns below "Similarity Range" show the number of predictions that are classified as similar according to each metric's thresholds, therefore, showing the number of true positives and false positives per metric. The columns below "Dissimilarity Range" follow a same protocol for the dissimilar predictions and list the number of true negatives and false negatives per metric. The "Uncertainty Range" part shows the predictions classified as "uncertain' according to each metric's thresholds.

Table 3 shows the precision, recall, and accuracy (the fraction of times the predicted similarity/dissmilarity values match the ground truth similarity/dissimilarity values) for each of the metrics based on the results of Table 2. Here, we considered any value below the similarity threshold (including the uncertainty region) to be a negative (dissimilar) prediction.

**RQ1: Is it possible to detect similar and dissimilar DNN functions using random inputs?** As the results show, the similarities and dissimilarities are pretty much distinguishable for all the three metrics. Visually, that can be seen in the large clusters of each plot on the right above the similarity threshold, and on the left below the dissimilarity threshold. The strong precision (over 96%), recall (over 76%), and accuracy (over 76%) numbers from Table 3 also attest this observation. Thus, we conclude that random inputs can be used for the purpose of DNN functions' similarity detection.

**RQ2: How do various similarity metrics compare?** We look at the last row of Table 2 to answer this question. We see that Spearman metric has the maximum number of true positive and true negative cases, and the least number of false positive and uncertain cases compared to the other two metrics. The overlap metric, on the other hand, has the least number of false negatives. These results, along with the precision, recall and accuracy numbers from Table 3 show that in general, Spearman is a powerful metric for detecting DNN functional similarities.

Looking at the individual cases, however, we see that there are scenarios where the metrics have different behaviors. One case is the *CNN vs. compat shape*: the CCA and the Spearman metrics were not able to detect any true positives in this case, while the Overlap metric detected 12 true positive cases. A distinguishing feature of this model is that it is a convolutional neural network model, while none of the models being compared to it (except for one) have a convolutional architecture. This shows that architectural differences between the models being compared may pose challenges for the similarity metrics and prevent them from making accurate predictions. We will fully study this issue in Section 5. For now, however, the results show that the Overlap metric, is still able to detect a few true positive cases in such scenarios, as opposed to the other two metrics that predicted almost all cases to be dissimilar.

These results show that in general, Spearman is a suitable metric for detecting DNN functions' similarity. In certain situations, for example where there is substantial differences between models' architectures (which can be checked given the model file), the overlap metric (with BRINC inputs) is a more suitable choice. When compared with Spearman, CCA is not showing superior results. It, however, has a better recall and accuracy compared to the Overlap metric, while Overlap has a better precision.

Here, we also discuss some interesting comparison cases:

**MN Rev Clr vs. same shape**: One interesting observation here pertains to the Spearman metric that has identified MN Rev Clr to have an inverse relationship with several models (negative correlation values), reflecting the inverse coloring of the training datasets (as described in Section 3.2.2, we created the MN Rev Color dataset by reversing the colors of the background and the digits in the MNIST dataset; that was to change the background color to black and the digits' color to white). The other observation is one model with accuracy > 90% on MN Rev Color's dataset and a high similarity value with this model using all three metrics. This case is circled in all scatter plots of Figure 4(b). Since MN Rev Color's training data was curated by us by reversing the colors, we did not expect to find any similar functions. So we investigated this case's repository and its training code, and it turned out that the same coloring transformation that we applied to the MNIST dataset to create the MN Rev color dataset is applied on this model's dataset too. This model is also circled in the Spearman scatter plot of Figure 4(a): the Spearman metric calculated a similarity of $\approx -0.25$ for this model and the MN Ver1 reference model, showing a moderate inverse relationship between these two.

**CNN vs. compat shape**: The odd case in this group is a model with very low accuracy (less than 20%) that scored above the similarity threshold only with the overlap metric, circled in the Overlap scatter plot of Figure 5(c). This could be evidence of a Type-1 error (false positive), so we analyzed this model's GitHub repository: this is a case of a model that performs digit recognition with digits similar to what exist in the MNIST dataset, but trained on a different dataset created by the author of that project. As such, the overlap metric with the use of BRINC generated input was able to flag it as similar to our CNN model. The other two metrics did not detect this case.

### 4.2 Unknown Models

Here we answer RQ3 about the possibility of finding similarities among unknown models. The purpose here is to understand how the studied metrics, with the use of random inputs, would work in cases where we do not have any information about the reference or the candidate model (in previous section, we had knowledge about the reference models as they were trained by us). To this aim, we randomly selected two clusters from our D56K dataset and one classifier from each cluster to serve as a reference model. Then, for each cluster, we calculated similarity scores between the selected reference model and all other models in the cluster, using the two metrics that in Section 4.1 we observed to have the highest precision: Spearman and Overlap.

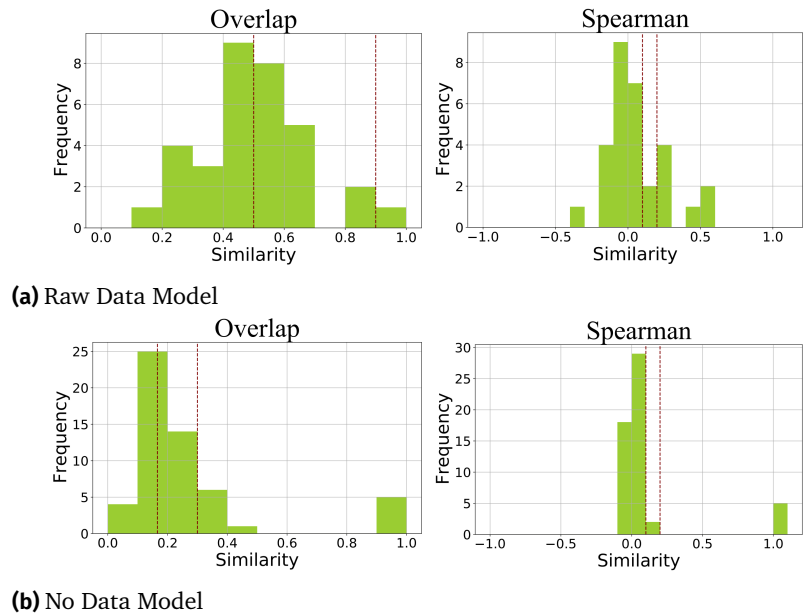

(a) Raw Data Model

(b) No Data Model

**Figure 6** Similarity values for arbitrary models

For one reference model, a raw dataset file with string and categorical variables was found in the GitHub repository along with a code transforming these feature values to numerical values appropriate for the DNN model. Therefore, although a dataset is available for this model, using it needs unknown processing steps and human effort that are hard to automate. In the case of the other reference model, we did not find any related dataset files in the GitHub repository. Figure 6 shows the similarity values when each cluster's reference model was compared with all other models in the cluster (x-axis) and the frequencies for these similarity values (y-axis), per metric. The dashed vertical lines show each metric's dissimilarity and similarity thresholds based on the thresholds discussed in Section 3.4.

**Reference Model with Raw Data Available:** The cluster including this reference model, in total includes 34 models for which the input is a vector of 26 and the output is binary. The reference model's GitHub repository describes it as aimed at "telecommunication customer churn detection". Based on Figure 6(a), Spearman detected 7 similar cases. From these, the Overlap metric detected one as similar, and placed the rest in the uncertainty region, two with a high similarity (over 0.8) and the other four with ≈ 0.61 similarity. For the three models with highest similarities by both metrics, their repositories indicate them to be performing churn prediction, hence, we considered them to be true similar cases. The repositories for the other four models describe them as aimed at people's distress recognition or analysis, and therefore, we did not consider them to be true positives by reported by Spearman.

**Reference Model with No Data Available:** The cluster including this model has 56 models with input shape of 48 × 48 × 1 and output shape of 6. The reference model's repository describes it as a "facial emotion recognition" project. As the plots in Figure 6(b) show, both metrics detected five exact matches (100 % similarity): in three cases, the models' repositories description mention face or emotion recognition, and in two cases, no information is given. The Overlap metric detected other similar cases too: one with $\approx 0.42$ similarity where the repository's description is emotion recognition and six cases with similarities $> 0.3$. In three cases of these six cases, the repository mentions either face or emotion recognition, and in the rest, no description was found. These cases were not detected by the Spearman metric.

Based on these results, we can answer **RQ3: Is it possible to detect similarities within unknown models using random inputs?** As attested by the results, we were able to detect similarities among the models for which we do not have any (or have little) information, with the use of random inputs.

### 4.3 Discussion

As the results presented in this section show, random inputs are a viable replacement of canonical test inputs for purposes of similarity detection of DNN classification functions using the the three studied metrics. Spearman correlation shows the best results in most experiments. Spearman is also easy to use, as it needs simple unconstrained random inputs, and produces results bound to the range of $[-1, +1]$ with well established dissimilarity/similarity thresholds. This metric is also able to detect inverse relationships between the functions. CCA is also easy to apply, and it is similar to Spearman correlation in terms of inputs and thresholds, and produces similarity results in the range $[0, 1]$. The overlap metric is more complicated to apply, as it requires the generation of balanced random inputs. However, this metric was shown to perform better when there are substantial architectural differences between models. In terms of number of inputs, we observed that in case of balanced random inputs, the number of needed random inputs correlates with the number of output classes; the more the number of outputs, more inputs are needed to be able cover all output labels at a sufficient quantity. For other two metrics, we observed that a number of inputs more than $3,000$ give satisfying results, however, the results reported in this paper were based on $20,000$ randomly generated inputs.

All in all, as the results show, the three metrics complement each other; there are cases where one metric detects similarity and the others fall within uncertainty ranges. Therefore, for the most accurate results, a combination of the metrics is recommended, and the final conclusion can be derived based on the results of all.

## 5 Sensitivity Analysis

Here, we present a set of experiments using Dataset D14 to answer RQ4 and RQ5. To answer these research questions, we present Figure 7 that shows the box&whiskers plot of the similarity (for each metric) between the reference model (MN Ver1) and all the other models of dataset D14. Each box corresponds to 10 different random input datasets (unconstrained inputs for CCA and Spearman, and BRINC inputs for Overlap). The parameter values to generate BRINC inputs were: $distance = 0.001, mutPer = 5\%, ranges = \{(-1, 0), (0, 1), (-1, 1)\}, maxMut = 300, maxValid = 1000$.

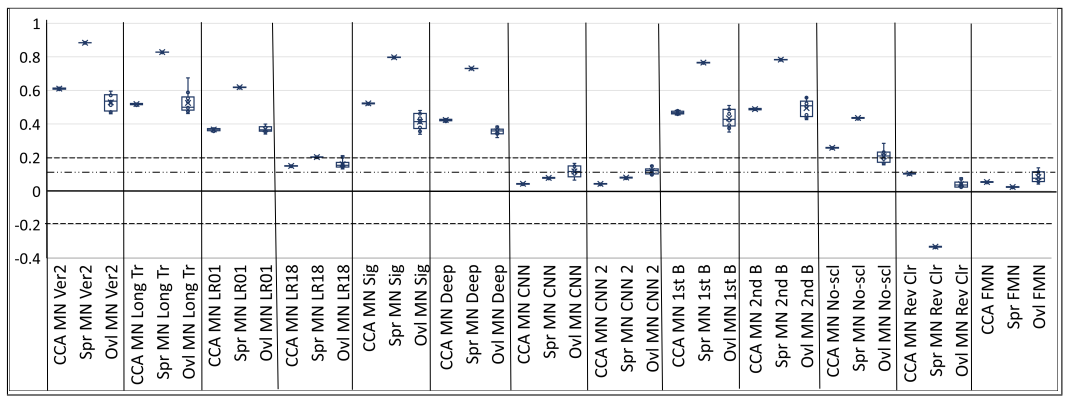

**Figure 7** Similarity with MN Ver1. Each box: min, $1^{st}$ quartile, median, $3^{rd}$ quartile, max.

**RQ4: What is the effect of input generation randomization on functions' response to similarity detection?** We look at the variations of similarity results when computed using 10 seperately generated random inputs to answer this question. Figure 7 shows that the effect of randomization is negligible for CCA and Spearman metrics. The boxes corresponding to these two metrics are small with minimum, maximum, and median being almost the same. For Overlap metric, we see some variability within the boxes, which is, however, bound to relatively narrow bands. That can be seen in the values of the $1^{st}$ and $3^{rd}$ percentiles, as well as in the min and max values for each box. In particular, we notice that the variability is lower for lower values of similarity, and that, for the higher values of similarity, the minimum is very far from the thresholds. This shows that one single run of input generation will often be enough to determine functional similarity. When the first measurement falls close to the thresholds, it may be necessary to repeat the measurements and take the median as the best approximation.

**RQ5: What is the effect of models' training related characteristics on their response to similarity detection?**
**Training randomization:** We look at the "Instances" group, MN Ver1 and MN Ver2, two virtually identical models with virtually identical accuracies. As Figure 7 shows, Spearman and Overlap predicted the highest median of similarity for them, and CCA predicted one of the highest, as we expected.
**Training parameters:** Looking at the models in "Train Params" group in Figure 7, we see that increasing the number of epochs (MN Long Tr) doesn't seem to have a strong effect on the measured similarity. The learning rate (MN LR01 and MN LR18), however, seems to have a measurable effect. For MN LR01, the similarity values are still well above the similarity threshold, but for MN LR18, the numbers are much lower. The reason may be that increasing the learning rate makes the models converge faster and this may hurt their generalization abilities.
**Models' accuracy:** Most of the studied models have accuracies > 95 %, with two exceptions: MN LR18 ($\approx$ 76 % accuracy), and FMN ($\approx$ 89 % accuracy). Figure 7 shows that the median value of MN LR18's similarity with the reference model is either close (Spearman) or inside (CCA and Overlap) the uncertainty region. For FMN model that has a better accuracy, we see that the median values are correctly below the dissimilarity threshold. This means that the similarity values are sensitive to the accuracy of the models if the accuracy is sufficiently low. The reason can be related to the fact that low accuracies hurt models' prediction capabilities.

**Architectures:** The Architectures group includes two models (MN Sig and MN Deep) with minor architectural differences (i.e., differences in the organization of the models' layers) with the reference model, and two models (MN CNN and MN CNN2) with substantially different architectures with the reference model, as they include convolutional layers. Figure 7 shows that for the models with minor differences (MN Sig and MN Deep), the medians of similarity are well above the similarity threshold. The similarity values for CNN models, however, are barely above the chance threshold for the overlap metric, and always in the dissimilarity region for the other two metrics. This shows that similarity detection with random inputs can fail to compute accurate similarity measurements when the architectures are substantially different.

**Training datasets:** Looking at the Datasets group in Figure 7 (MN 1st B onward), we see that similarity calculations are relatively robust with respect to functionality-preserving training data changes (MN 1st B, MN 2nd B, MN No-scl), and are correct in classifying as dissimilar the models trained on data that produces functionally different classifiers (MN Rev Clr and FMN).

## 6 Related Work

**Similarity of Neural Networks' Representations.** There is a line of work focused on the similarities of neural networks' representations learned at different layers. Laakso et al. [19] propose to compare neural networks' representations by comparing the distances among neural activations. Raghu et al. [28] propose to use Canonical Correlation Analysis (CCA) to measure the similarity between neural network representations. Based on that, Morcos et al. [23] propose projection weighted CCA. Kornblith et al. [17] discuss that CCA cannot measure meaningful similarities between the representations of higher than the number of datapoints and propose a similarity index, named CKA. Li et al. [20] study whether separately trained DNNs learn features that converge to similar spaces and Wang et al. [37] study whether neural networks that have identical architectures but are trained using different initializations learn similar representations. Unlike these work, we seek to find DNN similarities from a functional perspective, and in the absence of meaningful inputs.

**Fuzz Testing Neural Networks.** We are also inspired by the work on generating inputs for testing DNNs. DeepXplore [26] proposes neuron coverage as a test coverage metric. Later research [12] finds neuron coverage to be suffering from problems such as generating unnatural inputs and being biased towards certain labels. TensorFuzz [24] measures coverage based on the activations that are activated (typically logits or one layer before) and a new coverage is found if the distance of the vector of activations is greater than a threshold. DeepHunter [42] proposes a metamorphic mutation strategy that needs domain knowledge. DiffChaser [43] finds disagreements among multiple models (mainly a model and its variants) to help in model debugging.

## 7 Threats to Validity

In all experiments, we assumed that the ordering of output labels is the same for the reference and the candidate models. Although it is unlikely to have different orderings, specifically for well-known datasets, the existence of such cases can affect our results, both the calculated accuracy of the models and the similarity predictions. We used the mean value to aggregate the correlation values over models' output as it gives a holistic summarization over the calculated correlation values for all the output classes. Use of other statistical metrics for this purpose needs further investigation.

## 8 Conclusions and Future Work

In this paper, we formulated and discussed the problem of finding functional similarities among neural classifiers in the absence of canonical inputs, and studied three metrics that can be used for this purpose. To the best of our knowledge, this is the first large-scale study aimed at this problem. We also introduced a method, BRINC, for generating balanced random inputs for classifiers to be used with the Overlap metric. Our results show that similarity detection of DNN functions using random inputs is possible, and that all three studied metrics provide promising results, the Spearman metric, however, stood out, both in terms of ease of use and the overall results.

We aim to extend this work beyond classifiers, and investigate suitable metrics and inputs that can be used to detect similarities and dissimilarities among other kinds of neural functions. Also, another interesting line of related work is to investigate the applicability of the proposed similarity metrics using random inputs in scenarios of model theft, model search, and malware detection.

## A Appendix

The following sections detail the issues caused by using unbalanced inputs with the Overlap metric as well as a method for generating balanced random inputs to be used with the Overlap metric.

### A.1 Issues with Unbalanced Inputs and Overlap Metric

The two plots on top of Figure 8 illustrate the issue caused by using unbalanced inputs and the Overlap metric. These plots show the overlap similarity predictions (y-axis) for two similar digit recognition models against many others using the same set of unbalanced random inputs. The x-axis here shows the candidate models' accuracy on the reference model's testing dataset, therefore, being the ground truth of similarity. In spite of the two reference models being similar with each other, their similarity predictions with the other models are very different, and, even worse, are scattered, making it difficult to decide where to draw the threshold line to decide about the similarity. The situation is even worse with Model2 where the similarity measurements for both similar and dissimilar models reach ≈ 40%. Any chosen line can result in several false positives.

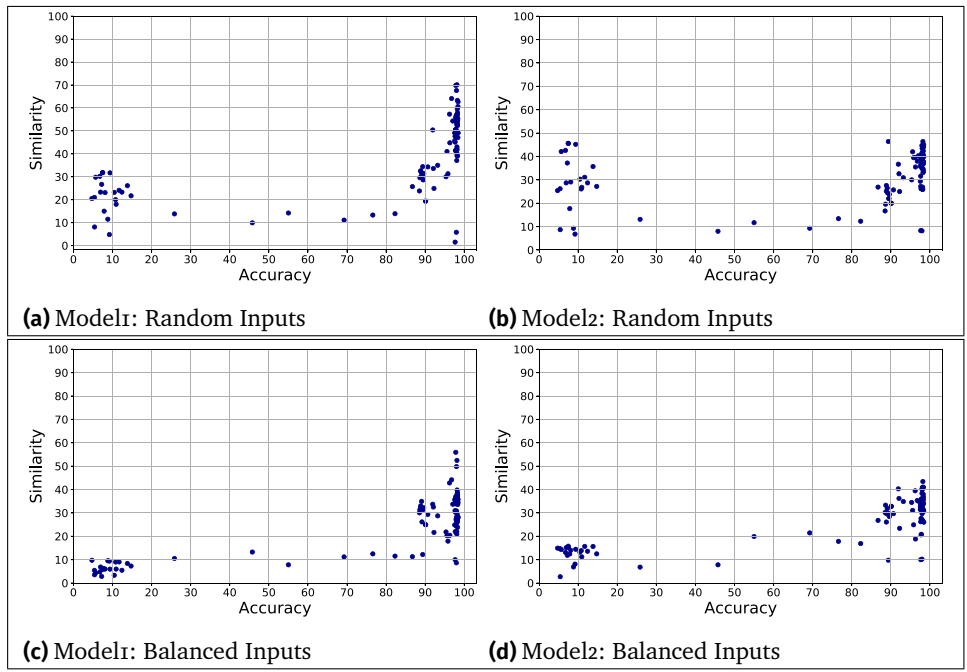

**Figure 8** Overlap metric: balanced vs unbalanced inputs

In practice, we observed that with unconstrained generation of random inputs, these inputs end up being biased towards a set of labels, therefore, endangering the specification of similarity thresholds for the Overlap metric. This motivates the next option for the input generation process: to impose some constraints so that the *reference model's predicted labels on the generated random inputs have a balanced distribution*. The two plots on the bottom of Figure 8 show the similarities for the same two models on balanced inputs. Here, the similarity values are clustered, and it is possible to establish a threshold that fit both cases without false positives (the similar models have predicted similarity values above ≈ 20%).

### A.2 Balanced Random Inputs Generation

We generate balanced random inputs using an algorithm that we call BRINC (Balanced Random Inputs for Neural Classifiers). The generated inputs are balanced with respect to the reference model's predictions, and therefore, can be used with the overlap metric to estimate DNN functions' similarity. The algorithm works as follows:

**Seed inputs.** BRINC starts by generating a few seed inputs that cover all output labels in a balanced format. Starting with a set of initial inputs (we included one with zero values, and one with random float values between -1 and 1), the seed inputs are repeatedly mutated by various mutation percentages to generate a few seed inputs that cover all the output labels.

**Next input to mutate.** After generating the seed inputs, at each step, one input is selected randomly from the set of inputs that predict the least frequent label at that point, to get mutated. This increases the chances that the mutant also predicts the same least frequent label, ensuring balance of the input dataset.

---

**Algorithm 1** Mutate an Input

---

1: **procedure** MUTATE(*input, range, mutPer, model*)
2:     *inputFlat* ← *getFlatShape(input)*
3:     *randIndexes* ← *randomSample(inputFlat, mutPer)*
4:     **for** each *index* ∈ *randIndexes* **do**
5:         *valueAtIndex* ← *randomFloatInRange(range)*
6:         *input[index]* ← *valueAtIndex*
7:     **end for**
8:     **return** *transformInputShapeToModelShape(input, model)*
9: **end procedure**

---

**Mutation.** The selected input is mutated to generate a new input. Algorithm 1 shows the steps: *mutPer* percent of the *input*'s values are changed to random float values. Parameter *range* is a tuple containing the range for generating these values. At the first step, the input is flattened (reshaped to a vector) to have a uniform process for mutation of inputs at any shape. Next, a set of indexes from this vector are randomly selected and changed. Finally, *input* is transformed to its original shape and returned.

**Mutant acceptance.** The mutant is added to the input corpus if (i) it adds new coverage to the network enforced by ensuring that the euclidean distance between the mutant's prediction probability vector and all other inputs' prediction vectors is more than a *distance* threshold, and, (ii) it predicts the least frequently predicted label (to ensure the balance of inputs). The first condition is an adaptation of measuring coverage as proposed by Odena et al. [24] where distances among logits are used to keep track of the coverage. It ensures that the prediction vectors of the generated inputs are distant (by some threshold) from each other so that the generation of many similar inputs is prevented.

**BRINC driver and parameters.** Algorithm 2 shows BRINC's input generation procedure and its parameters. The parameters are also summarized in Table 4. We tune the parameters based on the model's number of labels, the number of inputs that we would like to generate (controlled with *maxValid*), and the desired diversity (controlled with *distance*) in predictions. For the *ranges*, we chose the list $\{(-1,0),(0,1),(-1,1)\}$. We experimented with other ranges as well, but did not observe major differences. We also tried different values for *maxMut* and found the value 300 to work well. For all input generations, we started with the parameters suggested in Table 4 and further tuned the parameters if necessary.

■ **Table 4** Summary of BRINC's parameters.

| Param | Definition | Sugg. Value |
|---|---|---|
| *mutPer* | % of the *input*'s values changed at each step | 5% |
| *distance* | Min. Euclidean distance threshold among mutants | 0.001 |
| *ranges* | Set of ranges to constrain the numbers generation | $(-1,0),(0,1),(-1,1)$ |
| *maxMut* | Limit for the num. of mutations made in each range | 300 |
| *maxValid* | Max. num. of valid mutants generated in each range | 500 or 1000 |

**Algorithm 2** Generate Balanced Inputs

```
 1: procedure GENINPUT(model, distance, mutPer, ranges, maxMut, maxValid)
 2:     seeds ← generateSeeds(model, ranges)
 3:     for each range ∈ ranges do
 4:         generatedCount ← 0
 5:         noNewNum ← 0
 6:         while noNewNum ≤ maxMut and generatedCount ≤ maxValid do
 7:             nextToMutate, leastFreqLbl ← findLeastFreqLbl(seeds)
 8:             mutant ← doMutatate(nextToMutate, range, mutPer)
 9:             mutantLbl ← getLabel(model, mutant)
10:             distToNearest ← findNearestNeighByPred(mutant, seeds)
11:             if distToNearest > distance and mutantLbl == leastFreqLbl then
12:                 seeds.add(mutant)
13:                 generatedCount ← generatedCount + 1
14:                 noNewNum ← 0
15:             else
16:                 noNewNum ← noNewNum + 1
17:             end if
18:         end while
19:     end for
20:     return seeds
21: end procedure
```

## About the authors

**Farima Farmahinifarahani** is a PhD graduate in software engineering from the University of California, Irvine. Contact her at farimaf@uci.edu.

**Cristina V. Lopes** is a professor of software engineering at the University of California, Irvine. Contact her at lopes@uci.edu.